# Synthesis of multilamellar walls vesicles polyelectrolyte-surfactant complexes from pH-stimulated phase transition using microbial biosurfactants

**Chloé Seyrig[a], Patrick Le Griel[a], Nathan Cowieson,[b] Javier Perez,[c] Niki Baccile[a]**

[a] Sorbonne Université, Centre National de la Recherche Scientifique, Laboratoire de Chimie de la Matière Condensée de Paris , LCMCP, F-75005 Paris, France

[b] Diamond Light Source Ltd, Diamond House, Harwell Science & Innovation Campus, Didcot, Oxfordshire, OX11 0DE

[c] Synchrotron SOLEIL, L'Orme des Merisiers Saint-Aubin, BP 48 91192 Gif-sur-Yvette Cedex

**Abstract**

Multilamellar wall vesicles (MLWV) are an interesting class of polyelectrolyte-surfactant complexes (PESCs) for wide applications ranging from house-care to biomedical products. If MLWV are generally obtained by a polyelectrolyte-driven vesicle agglutination under pseudo-equilibrium conditions, the resulting phase is often a mixture of more than one structure. In this work, we show that MLWV can be massively and reproductively prepared from a recently developed method involving a pH-stimulated phase transition from a complex coacervate phase (*Co*). We employ a biobased pH-sensitive microbial glucolipid biosurfactant in the presence of a natural, or synthetic, polyamine (chitosan, poly-L-Lysine, polyethylene imine, polyallylamine). *In situ* small angle X-ray scattering (SAXS) and cryogenic transmission electron microscopy (cryo-TEM) show a systematic isostructural and isodimensional transition from the *Co* to the *MLWV* phase, while optical microscopy under polarized light experiments and cryo-TEM reveal a massive, virtually quantitative, presence of MLWV. Finally, the multilamellar wall structure is not perturbed by filtration and sonication, two typical methods employed to control size distribution in vesicles. In summary, this work highlights a new, robust, non-equilibrium phase-change method to develop biobased multilamellar wall vesicles, promising soft colloids with applications in the field of personal care, cosmetics and pharmaceutics among many others.

**Keywords**. Polyelectrolyte-Surfactant Complex, complex coacervates, biosurfactants, polyelectrolytes, multilamellar walls vesicles

**Introduction**

Polyelectrolytes and surfactants may assemble into complex structures known as polyelectrolyte-surfactant complexes (PESCs). When these compounds are oppositely charged, their self-assembly process is mainly driven by electrostatic interactions and it results in the formation of aggregates, which have a broad range of applications in biological materials,[1–5] drug delivery,[6–8] surface modifications,[9] colloid stabilization[10] and flocculation[11] and consumer health-care products. The rich mesoscopical and structural organisation of surfactants combined with the electrostatic interactions with polyelectrolytes give rise to a wide range of structures and phases.[12–18] Many works reported cubic or hexagonal mesophases[15,16] but also a number of micellar-based structures: pearl-necklace morphologies,[2,19,20] interpenetrated polyelectrolyte-wormlike/cylindrical micelles network,[19,21–23] spheroidal clusters composed of densely packed micelles held by the polyelectrolyte, the latter known as complex coacervates (*Co*) when they form a liquid-liquid phase separation.[19,24,25]

Very interesting PESCs structures are formed when the surfactant forms low curvature vesicular morphologies. It is in fact generally admitted that modifying vesicles by the addition of polyelectrolytes is an interesting, cheap and simple approach to obtain nanocapsules,[23] which are good candidates to be used as versatile delivery systems,[19,23] like gene delivery,[1,22,26,27] or as MRI contrast agents.[28] One of the first PESCs vesicular systems has been reported more than 20 years ago in DNA-CTAB (cetytrimethylammonium bromide) systems, which were the precursors of a number of carriers for gene transfection and often referred to as lipoplexes, when cationic lipids replace surfactants in DNA complexation.[29,30] If the term lipoplex supposes the use of nucleic acids as complexing agents, similar structures, often addressed to as onion-like structures[31] or multilamellar vesicles,[13] were observed using both lipids and surfactants complexed by a wide range of polyelectrolytes. However, multilamellar, or onion-like, vesicles are rather characterized by single-wall membranes concentrically distributed from the outer to the inner core of the vesicle. Lipoplexes, on the contrary, are vesicular objects with a large lumen and a dense multilamellar wall. For this reason, in this work we employ the name multilamellar wall vesicles (MLWV).

The mechanism of formation of MLWV was addressed by several authors, but a common agreement is not achieved, yet. Several works propose that the lipid:polyelectrolyte ratio controls the fusion of single-wall vesicles into MLWV,[19,29,32–35] while others rather observe vesicular agglutination under similar conditions.[36–38] In fact, a general consensus has not been found and a multiphasic system including agglutinated vesicles and MLWV are actually observed.[39] The question whether or not MLWV, and PESCs in general, are equilibrium structures and how they are formed is still open, especially when they are prepared under non-equilibrium conditions.[19] To the best of our knowledge, the only works exploring a stimuli-induced approach in the synthesis of MLWV in particular, and PESCs in general, were proposed by Chiappisi *et al.*.[21,40] However, the pH variation in these works was still performed under pseudo-equilibrium conditions with equilibration times ranging from 2 to 15 days for each pH value.

In a recent work, we have explored a *Co*-to-*MLWV* phase transition under non-equilibrium conditions using a continuous variation in pH,[41] as illustrated by Figure 1. We could show that in the presence of G-C18:1, an acidic microbial glycolipid biosurfactant,[42,43] and poly-L-lysine (PLL), a cationic polyelectrolyte (PEC), the pH-stimulated micelle-to-vesicle phase transition of the lipid drives a continuous, isostructural and isodimensional, transition between complex coacervates and MLWV. PLL strongly binds to the lipid monolayers thus favouring ($\Delta G$= -36.4 ± 1.9 kJ/mol) the formation of the multilamellar wall through both specific ($\Delta H$= -2.8 ± 0.8 kJ/mol, electrostatic and possibly hydrogen bonding) and non-specific ($\Delta H$= 28.9 ± 0.9 kJ/mol, entropic, hydrophobic effect) interactions, as quantified by isothermal titration calorimetry experiments.[41]

In the present work, we generalize the method of preparing MLWV through a phase transition approach performed under non-equilibrium conditions and we show its performance in comparison to the more accepted method of vesicular agglutination. We show that the former can be applied to a broader set of polyelectrolytes and we explore in more detail the structure and size control of MLWV.

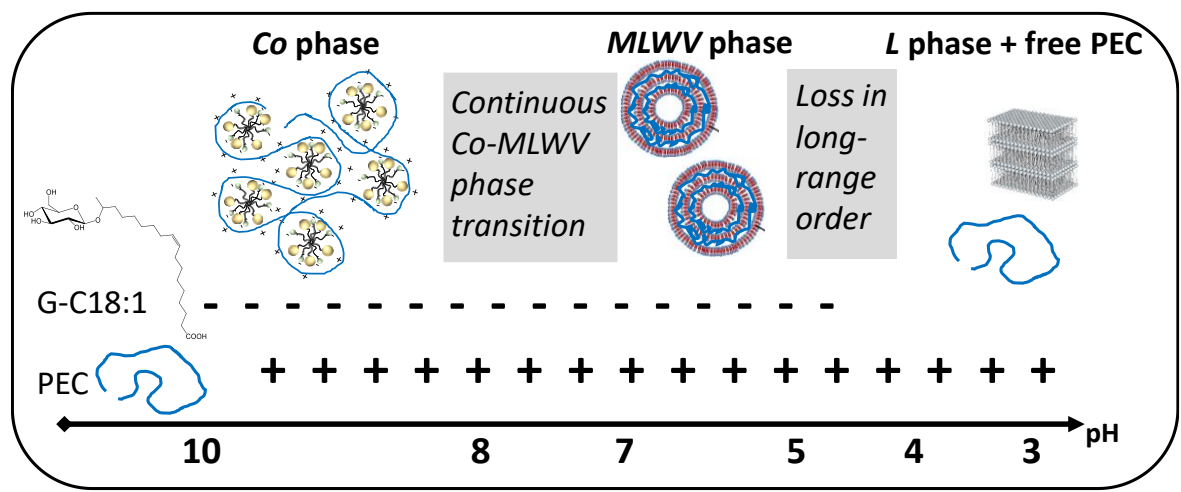

**Figure 1 – Phase transition and structures obtained by mixing G-C18:1 and PEC (chitosan, poly-L-lysine, polyallylamine or polyethylenimine) upon a rapid variation of pH. G-C18:1 is negatively charged between about pH 4 and alkaline pH, while PECs are positively charged below pH ~10 (depending on the exact pKa, given in the materials and method section). Complex coacervates (*Co*) composed of G-C18:1 and PEC form**

at pH ~10. They progressively rearrange into MLWV and dissociate below pH ~4, where G-C18:1 forms a lamellar (*L*) phase coexisting with free polymer chains.[41]

**Experimental section**

**Chemicals**

In this work we use microbial glycolipids G-C18:1, made of a single *β*-D-glucose hydrophilic headgroup and a C18 fatty acid tail (monounsaturation in position 9,10). From alkaline to acidic pH, the former undergoes a micelle-to-vesicle phase transition.[42] Syntheses of glucolipid G-C18:1 are described in Ref [44] and [43], where the typical $^1$H NMR spectra and HPLC chromatograms are given. The compound used in this work have a molecular purity of more than 95%.

The polyelectrolytes used in this work are chitosan, obtained from the deacetylation of chitin from crusteans' shells, poly-L-lysine, widely used in biomedical field, polyallylamine and polyethylenimine. Chitosan oligosaccharide lactate (CHL) ($M_w$ ≈ 5 KDa, p$K_a$ ~6.5)[45] with a deacetylation degree >90%, poly-L-lysine (PLL) hydrobromide ($M_w$ ≈ 1-5 KDa, p$K_a$ ~10-10.5)[46] and polyallyllamine hydrochloride (PAH) ($M_w$ ≈ 1-5 KDa, p$K_a$ ~9.5),[46] polyethylenimine (PEI) hydrochloride (linear, $M_w$ ≈ 4 KDa, p$K_a$ ~8)[47] are purchased from Sigma-Aldrich. We also employ a polyampholite, gelatin (Aldrich, type A, from porcine skin, $M_w$ ≈ 50-100 KDa, isoelectric point 7-9), as a control. All other chemicals are of reagent grade and are used without further purification.

**Preparation of stock solutions**

G-C18:1 (*C*= 5 mg·mL$^{-1}$, *C*= 20 mg·mL$^{-1}$), CHL (*C*= 2 mg·mL$^{-1}$), PLL (*C*= 5 mg·mL$^{-1}$, *C*= 20 mg·mL$^{-1}$), PEI (*C*= 5 mg·mL$^{-1}$), PAH (*C*= 2 mg·mL$^{-1}$) and gelatin (*C*= 5 mg·mL$^{-1}$) stock solutions (*V*= 10 mL) are prepared by dispersing the appropriate amount of each compound in the corresponding amount of Milli-Q-grade water. The solutions are stirred at room temperature (*T*= 23 ± 2 °C) and the final pH is increased to 11 by adding a few µL of NaOH (*C*= 0.5 M or *C*= 1 M).

**Preparation of samples**

Samples are prepared by mixing appropriate volume ratios of G-C18:1 stock solutions at pH 11 and cationic polyelectrolyte (PEC) stock solutions, as defined in Table 1. The final total volume is generally set to *V*= 1 mL or *V*= 2 mL, the solution pH is about 11 and the final concentrations are given in Table 1. The pH of the mixed lipid-PEC solution is eventually decreased by the addition of 1-10 µL of a HCl solution at $C$= 0.5 M or $C$= 1 M. The rate at which pH is changed is generally not controlled although it is in the order of several µL·min$^{-1}$. Differently than in other systems,[48,49] we did not observe unexpected effects on the PESC structure to justify a tight control over the pH change rate.

**Table 1 - Relative volumes of G-C18:1 and PEC solutions to mix to obtain given concentrations**

| Volume | | | Concentration | |
|---|---|---|---|---|
| G-C18:1 stock solution / mL | PEC stock solution / mL | Water / mL | $C_{\text{G-C18:1}}$ / mg·mL$^{-1}$ | $C_{\text{PEC}}$ / mg·mL$^{-1}$ |
| 0.5 | 0.5 | 0 | 2.5 or 10 | 2.5 or 10 |
| | 0.25 | 0.25 | 2.5 | 1.25 |
| | 0.125 | 0.375 | 2.5 | 0.625 |

**Dynamic light scattering measurements (DLS)**

DLS experiments are performed using a Malvern Zetasizer Nano ZS90 (Malvern Instruments Ltd, Worcestershire, UK) equipped with a 4 mW He–Ne laser at a wavelength of 633 nm. Measurements were made at 25 °C with a fixed angle of 90° and three acquisitions per sample.

**pH-resolved *in situ* Small angle X-ray scattering (SAXS)**

*In situ* SAXS experiments during pH variation are performed at room temperature on two different beamlines. The B21 beamline at Diamond Light Source Synchrotron (Harwell, England) is employed using an energy of $E$= 13.1 keV and a fixed sample-to-detector (Eiger 4M) distance of 2.69 m. The Swing beamline at Soleil Synchrotron (Saint-Aubin, France) is employed using an energy of $E$= 12 keV and a fixed sample-to-detector (Eiger X 4M) distance of 1.995 m. For all experiments: the $q$-range is calibrated to be contained between ~$5\cdot10^{-3}$ < $q$/Å$^{-1}$ < ~$4.5\cdot10^{-1}$; raw data collected on the 2D detector are integrated azimuthally using the in-house software provided at the beamline and so to obtain the typical scattered intensity $I(q)$ profile, with $q$ being the wavevector ($q = 4\pi \sin\theta / \lambda$, where $2\theta$ is the scattering angle and $\lambda$ is the wavelength). Defective pixels and beam stop shadow are systematically masked before azimuthal integration. Absolute intensity units are determined by measuring the scattering signal of water ($I_{q=0}$= 0.0163 cm$^{-1}$).

The same sample experimental setup is employed on both beamlines: the sample solution ($V$= 1 mL) with the lipid and PEC at their final concentration and pH ~11 is contained in an external beaker under stirring. The solution is continuously flushed through a 1 mm glass capillary using an external peristaltic pump. The pH of the solution in the beaker is changed using an interfaced push syringe, injecting microliter amounts of a 0.5 M HCl solution. pH is measured using a micro electrode (Mettler-Toledo) and the value of pH is monitored live and manually recorded from the control room via a network camera pointing at the pH-meter located next to the beaker in the experimental hutch. Considering the fast pH change kinetics, the error on the pH value is ± 0.2.

**Polarized Light Microscopy (PLM)**

PLM experiments are performed with a transmission Zeiss AxioImager A2 POL optical microscope. A drop of the given sample solution is deposited on a slide covered with a cover slip. The microscope is equipped with a polarized light source, crossed polarizers and an AxioCam CCD camera.

**Cryogenic transmission electron microscopy (cryo-TEM)**

Cryo-TEM experiments are carried out on an FEI Tecnai 120 twin microscope operated at 120 kV and equipped with a Gatan Orius CCD numeric camera. The sample holder is a Gatan Cryoholder (Gatan 626DH, Gatan). Digital Micrograph software is used for image acquisition. Cryofixation is done using a homemade cryofixation device. The solutions are deposited on a glow-discharged holey carbon coated TEM copper grid (Quantifoil R2/2, Germany). Excess solution is removed and the grid is immediately plunged into liquid ethane at -180°C before transferring them into liquid nitrogen. All grids are kept at liquid nitrogen temperature throughout all experimentation. Images were analyzed using Fiji software, available free of charge at the developer's website.[50]

**Results**

In recent publications,[41,51] we have explored the complex coacervation between microbial glycolipids and PEC. For this reason, this aspect is only briefly shown here. Cryo-TEM images presented in Figure 2 show the structure of PEC-complexed G-C18:1 complex coacervates above pH 7. Above this value, we expect the G-C18:1 to be negatively charged and PEC positively charged, whereas the apparent p$K_a$ of G-C18:1 is expected to be between 6 and 7, similarly to oleic acid,[52] and the p$K_a$ of PECs being provided in the materials and methods section. Irrespective of the selected PEC, all systems show spheroidal colloids of variable size in the 100 nm range. One can identify two types of structures, both typical of complex coacervates:[24,25,51,53] dense aggregated structures, shown in Figure 2a,c and very similar to what was found by us[41,51] and others,[24] are attributed to dehydrated, densely-packed, micelles tightly interacting with the polyelectrolyte; a biphasic medium composed of spheroidal, poorly-contrasted, polymer-rich, colloids embedded in a textured, surfactant-rich, medium. The latter were also reported by us[41,51] and others.[53,54] In all cases, *Co* phase is a PESC forming in the micellar region of the surfactant's phase diagram and having the specificity of a liquid-liquid phase separation,[19,25] compared to other supramicellar PESCs undergoing a solid-liquid phase separation.[19]

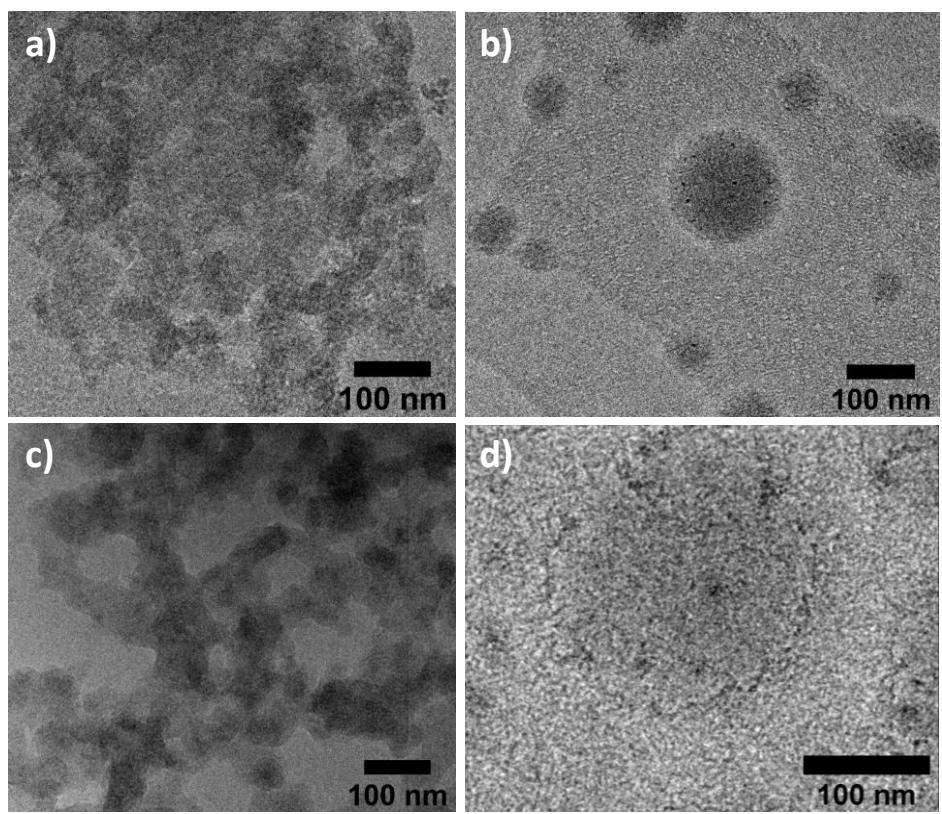

**Figure 2** – Cryo-TEM images of PESC solutions in the complex coacervate phase composed of G-C18:1 lipid complexed with a) CHL (pH 7.16), b) PLL (pH 9.16), c) PAH (pH 8.96) and d) PEI (pH 9.02). $C_{G\text{-}C18:1}$= $C_{PEI}$= 2.5 mg·mL$^{-1}$, $C_{CHL}$= 1 mg·mL$^{-1}$, $C_{PAH}$= 0.25 mg·mL$^{-1}$, $C_{PLL}$= 1.25 mg·mL$^{-1}$

The difference between dense and poorly-contrasted structures is PEC-independent and it is more related to the stage of coacervation. At an early stage, colloids with a relatively low electron density form and coexist with a rich micellar phase. Free micelles progressively interact with residual polymer chains. At a later, entropy-driven (dehydration and counterion release),[55] stage of coacervation, droplets with a higher electron density massively form. Unfortunately, neither the texture of the particles nor their internal structure can be easily controlled as they strongly depend on the type of PEC, its stiffness, charge density, stage of coacervation and even kinetics. For these reasons, isolating a specific structure in a *Co* phase can be challenging and we have ourselves found coexisting dense and poorly-contrasted structures,[41] thus preventing any reasonable structure-composition generalization concerning the images presented in Figure 2.

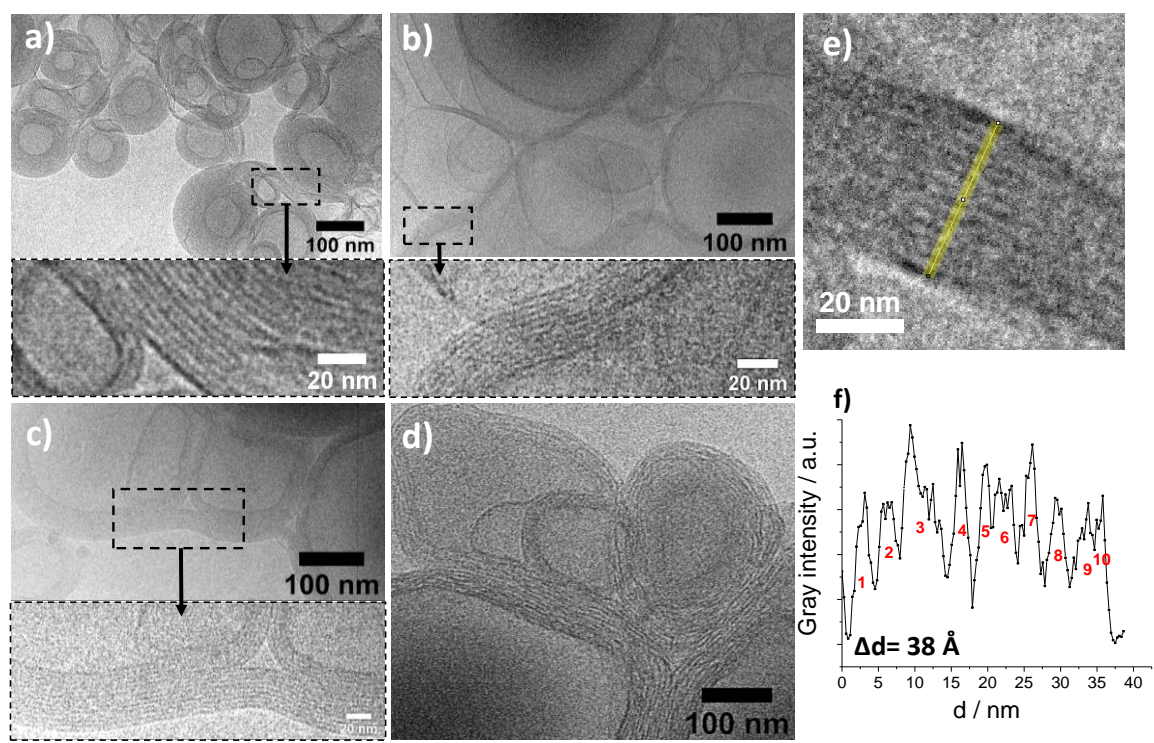

**Figure 3 – Cryo-TEM images of a *MLWV* phase composed of acidic G-C18:1 lipid complexed with a) CHL (pH 4.87), b) PLL (pH 4.70), c) PAH (pH 4.25) and d) PEI (pH 5.33). $C_{G-C18:1}= C_{PEI}= C_{PLL}= 2.5$ mg·mL$^{-1}$, $C_{CHL}= 1$ mg·mL$^{-1}$, $C_{PAH}= 0.25$ mg·mL$^{-1}$. e) Zoomed cryo-TEM image of [G-C18:1 + PLL] mixture and its corresponding profile (f) allowing the determination of the interlamellar distance. Cryo-TEM data have been analyzed using Fiji software.**[50]

At pH below 7, vesicular structures with multilamellar walls (*MLWV* phase) are observed by cryo-TEM for all PEC samples (Figure 3 and Figure S 1). These structures are closely-related to a lipoplex-type phase rather than to an onion-like phase: the latter is composed of concentric single-wall vesicles, while the former keeps a free lumen and a thick multilamellar wall.[29] Figure 3 also shows a strong packing of the multilamellar walls as well as a strong interconnection between adjacent vesicular objects, in agreement with lipoplexes and other MLWV reported in the literature.[23] The walls are constituted of alternating sandwiched layers composed of tightly packed polyelectrolyte chains and interdigitated monolayers of G-C18:1.[41] *d*-spacing can be directly estimated from cryo-TEM images (Figure 3e,f) and we find a set of values of *d*= 33.7 ± 4.95 Å for the PLL system and *d*= 31.6 ± 3.00 Å, 25.3 ± 4.60 Å and 41.1 ± 0.30 Å respectively for CHL, PAH and PEI systems. Within the error, these values are compatible with interdigitated G-C18:1 layers,[41–43] of which the thickness can be estimated to be about 25 Å by applying the Tanford relationship,[56] but also close to what is classically recorded for lipoplexes.[22,23,33] One may note that the multilamellar walls of the PECSs involving PEI (Figure 3d) appear more disordered than for other PESCs. At the moment, we do not have a clear explanation for that and we actually believe it to be an artifact due to freezing, because the full width at half maximum of the corresponding lamellar peak in SAXS experiments is $\Delta q$ ~$2 \cdot 10^{-3}$ Å$^{-1}$, the same value found for the PLL system.

Cryo-TEM images recorded on the *Co* (Figure 2) and *MLWV* (Figure 3) phases show that the *Co*-to-*MLWV* transition is a general property of G-C18:1 PESCs: it strictly depends on the lipid phase behavior, while the polyelectrolyte only guarantees the cohesion between the lipid membranes. We highlighted elsewhere[41] by pH-resolved *in situ* SAXS experiments an explicit isostructural and isodimensional continuity in the *Co*-to-*MLWV* phase transition: the broad correlation peak at $q$= 0.171 Å$^{-1}$ (*d*-spacing of 36.7 Å) of the coacervate phase coexists with the sharp diffraction peak of the *MLWV* phase at $q_1$= 0.178 Å$^{-1}$ (*d*-spacing of 35.3 Å) in a narrow range around pH 7.[41] Restructuring is driven by the progressive hydrogenation of the carboxylate group and the resulting conformational change of the lipid, which favors the formation of low curvature colloids, while inter-lipid repulsive electrostatic interactions disappear in the meanwhile.

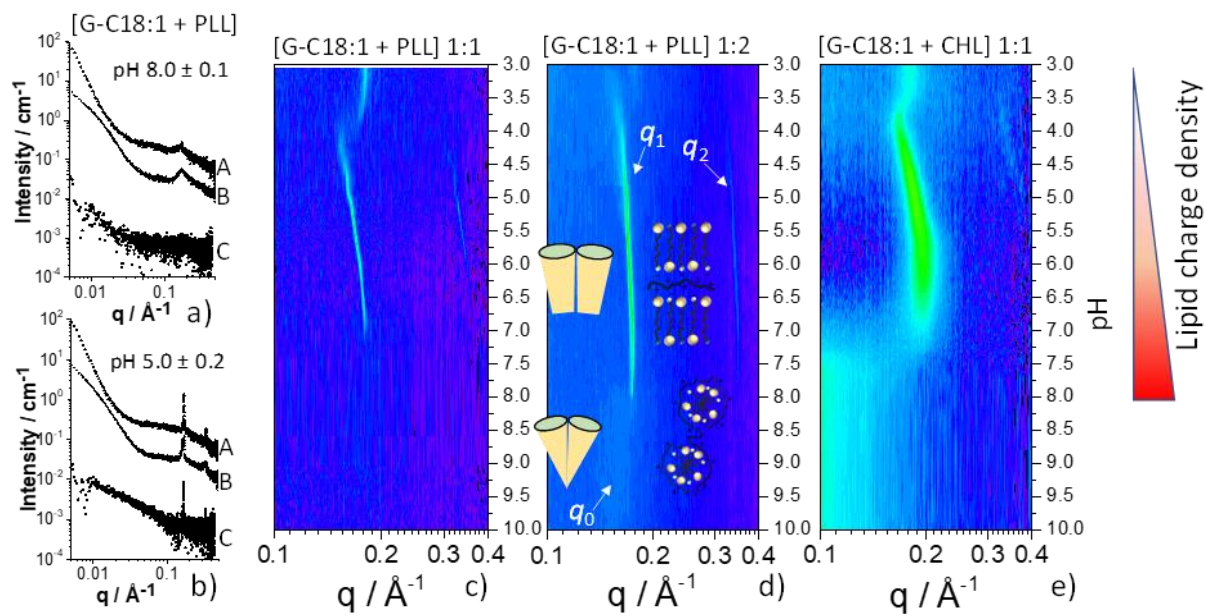

**Figure 4** - SAXS profiles of [G-C18:1 + PLL] PESCs at a) basic and b) acidic pH with G-C18:1:PLL concentration ratios in mg·mL$^{-1}$: A= 2.5:5, B= 10:10, C= 2.5:2.5. c-e) 2D SAXS contour plots of G-C18:1:PLL concentration ratios in mg.mL$^{-1}$: c) 2.5:2.5, d) 2.5:5 and e) 2.5:2. pH is varied from basic to acidic.

SAXS profiles presented Figure 4 show two different behaviors of the mixture [G-C18:1 + PLL]: at basic pH (Figure 4a), a broad correlation peak is observed at about $q= 0.17$ Å$^{-1}$ for all lipid:PLL ratios, where the peak can be more pronounced either with concentration (B profile) or lipid:PLL ratio (A profile). SAXS profiles B and C were previously assigned to complex coacervates, and more details on the structure of the *Co* phase can be found in Ref. [41]. In similar systems, the slope at low $q$ was shown to be indicative of the shape of the PESC;[40] here, the slope is below -3. If such values are typical of fractal interfaces,[57,58] we cannot unfortunately draw any conclusion on the structure of the complex coacervates, most likely because the *Co* phase in these systems is heterogeneous.[41]

Below pH 7 (Figure 4b), a sharp diffraction peak and its first harmonics are visible respectively around $q_1= 0.17$ Å$^{-1}$ and $q_2= 0.34$ Å$^{-1}$, characteristic of the (100) and (200) reflections of a lamellar order in the walls, described previously and shown in Figure 3. The *d*-spacing of 37 Å is in agreement with the ones deduced from cryo-TEM (Figure 3e,f). Similar results are obtained at different lipid:PLL ratios (Figure 4c,d) but also for other PEC. Figure S 2 presents the SAXS signals of [G-C18:1 + CHL] solutions at basic and acidic pH, compared to the control solutions of [G-C18:1] and [CHL] alone as well as their arithmetic sum. If at acidic pH the signature of the lamellar wall of the mixture compared to the controls is out of doubt, the signal at basic pH is less straightforward to interpret, due to the scattering of CHL alone, known to precipitate above pH 7.[59] This result is similar to what was found for acidic deacetylated sophorolipids;[41,51] however, considering the fact that cryo-TEM experiments suggest the presence of complex coacervates, we cannot exclude their formation, although their content may constitute a small minority, if compared to the PLL-based PESCs in the same pH range. Another argument for the formation of *Co* in the presence of CHL will be given later.

Figure 4c-e show the pH-resolved *in situ* contour plots of [G-C18:1 + PLL] PESCs at various lipid:PLL ratio and with CHL. They are recorded between pH 10 and 3 and focus on the high-$q$ region of the SAXS pattern, sensitive to the structural *Co*-to-*MLWV* phase transition. All pH-resolved *in situ* contour plots in Figure 4 show three common features: 1) the *Co*-to-*MLWV* transition between pH 8 and 7, where $q_1$ and $q_2$ refer to the first and second order peaks of the lamellar wall; 2) a low-$q$ shift of $q_1$ and $q_2$ when pH decreases to about 4.5, indicating a swelling of the lamellar period, and 3) a loss of the signal between about pH 4.5 and pH 3.5, below which a constant peak at higher $q$-values (generally around $q=0.2$ Å$^{-1}$) appears. These phenomena were quantitatively described in more detail in Ref. [41] and will only be summarized hereafter.

When fully deprotonated at basic pH, G-C18:1 is in a high curvature, micellar, environment (*Co* phase). This state, represented by the drawing superimposed on Figure 4d, is proven by both cryo-TEM and the broad correlation peak at about $q_0=0.17$ Å$^{-1}$. After crossing the transition pH range between 8 and 7, the number of negative charges decreases and G-C18:1 is entrapped in a low-curvature, interdigitated layer, environment. The continuity between $q_0$ and $q_1$ strongly suggests an isostructural and isodimensional transition between the micelle and membrane configurations, without any loss of the interaction with the polyelectrolyte. This is also sketched on Figure 4d. When the pH is decreased further, the COOH content increases and thus the membrane charge density decreases. The interlamellar distance consequently increases due to the repulsive pressure exerted by the charged polyelectrolyte, which undergoes hydration and increase internal electrostatic repulsion.[2,60,61] When hydrogenation of carboxylate groups reach a certain extent, attractive interaction with PLL can no longer hold the membranes together and MLWV then lose their long-range lamellar order, which results in their complete disruption and the concomitant expulsion of PLL. Below pH 3, this mechanism leads to the precipitation of a polyelectrolyte-free lamellar, *L*, phase, which is also observed for PEC-free G-C18:1 solutions.[41]

A closer look at the experiments in Figure 4 indicates two additional features. The pH stability domain of the *MLWV* phase seems to vary with the lipid:PLL ratio. Comparison of Figure 4c and Figure 4d, respectively recorded at lipid:PLL= 1:1 and 1:2 reveal that the $q_1$ peak of the *MLWV* phase is observed between pH 8 and 7. At the 1:2 ratio the *MLWV* phase starts at about pH 8 while at the 1:1 ratio the *MLWV* phase is only visible at pH below 7. At higher concentrations ($C$= 10 mg·mL$^{-1}$), but still for a 1:1 ratio, the stability frontier seems to be shifted at pH of about 7.5.[41] Although we do not have enough data to draw a general trend, it is well-known that the lipid:polyelectrolyte ratio reflects the negative:positive charge ratio and for this reason it has a direct impact on the electroneutrality, thus affecting a number of structural features of PESCs: the wall thickness of the multilamellar structure,[21,62] the PESC morphology and colloidal stability.[19] For instance, order is noticeably improved when the charge ratio approaches 1:1,[63] and micelle-polyelectrolyte complex coacervation can be favored or not.[64] This ratio is particularly crucial to control the properties of the lipoplexes and thus their applications: lipid/DNA ratio was reported to influence both the formation of lipoplexes and the release of DNA[65] and gene transfer activity.[66] Many authors have shown that the lipid:polyelectrolyte ratio actually controls the formation of MLWV structures[19,29,32–35] over agglutinated single-wall vesicles,[36–38] but in fact it is more likely that a general consensus has not been found, yet, and reality often consists in a mixtures of MLWV and agglutinated vesicles,[39] although many authors do not specify it. One of the reasons that could explain such discrepancy is the parallel influence of several other parameters like the charge density on both the lipid membrane and in the polyelectrolytes, the rigidity of the latter, the bending energy of the lipid membrane, the ionic strength and so on.[14,19] In the present case, it is important to note that: 1) G-C18:1 forms a stable *MLWV* phase with all PEC tested in this work and of different origin (biobased vs. synthetic) and rigidity. 2) MLWV are stable in the neutral pH range, which can be a good opportunity for applications in the biomedical field, for instance.

An interesting remark concerns the long-range order inside the vesicular multilamellar walls. The width of the lamellar peak around $q$ ~0.2 Å$^{-1}$ is more than ten times larger for the CHL (Figure 4e, $\Delta q$ ~3.10$^{-2}$ Å$^{-1}$) than the PLL (Figure 4c,d, $\Delta q$ ~2.10$^{-3}$ Å$^{-1}$) system, either suggesting an average smaller size of the lamellar domains or a poorer lamellar order in the case of the MLWV obtained from CHL. The reason behind such difference could be the bulkiness and stiffness of CHL with respect to PLL,[32] but one should recall from Figure 2 and related discussion that [G-C18:1 + CHL] solutions do not form an extensive *Co* phase. We have already made the hypothesis that the *Co* phase is necessary to form the *MLWV* phase,[41] and we will reinforce this assumption in the next part of this work.

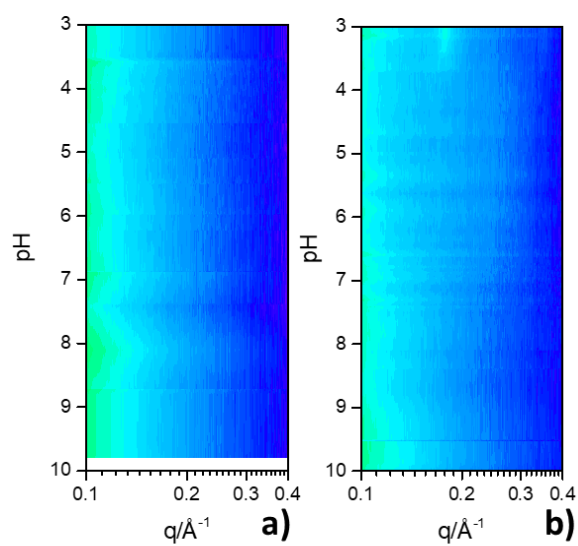

**Figure 5 – pH-resolved *in situ* 2D SAXS contour plots of a) gelatin ($C$= 2.5 mg.mL$^{-1}$) and b) [G-C18:1 + gelatin] mixture ($C_{\text{G-C18:1}}$= $C_{\text{Gelatin}}$= 2.5 mg.mL$^{-1}$).**

The data collected so far show that G-C18:1 interacts with all polyelectrolytes tested in this work and that its micelle-to-vesicle phase transition drives the *Co*-to-*MLWV* transition. As one could reasonably expect, and actually confirmed by ITC,[41] strong specific electrostatic interactions, electrostatic n nature, between the positively charged PEC and negatively charged G-C18:1 drive the PESC formation across the entire pH range. To test the solidity of the PESCs synthesis using G-C18:1 and PECs, we employ gelatin, a polyampholyte, as a possible alternative to polyelectrolytes and which could be interesting to prepare biobased PESCs. We use a commercial (Aldrich) source of gelatin type A, a natural protein of isoelectric point between 7.0-9.0, below which the charge becomes positive. Figure 5 shows pH-resolved *in situ* contour plots of gelatin and [G-C18:1 + gelatin] samples. The control gelatin sample in Figure 5a shows no specific contribution across the entire pH range between $0.1 < q\ /\ \text{Å}^{-1} < 0.4$. Interestingly, the [G-C18:1 + gelatin] sample presented in Figure 5b does not show any signal either in the same pH and *q* range, except for the systematic signal of the lamellar, *L*, phase of G-C18:1 below pH 4.[41,42]

Despite an expected positive charge density of gelatin, the *in situ* SAXS experiment shows no sign of the *Co* phase above pH 7, indicating that the charge density is probably too low to interact with negatively charged G-C18:1 micelles. Although somewhat unexpected because interactions with negatively charged sodium dodecyl sulfate micelles across a wide compositional and pH range were reported in other studies,[67] this result is not a surprise. What it is more interesting from a mechanistic point of view is the lack of the *MLWV* phase below pH 7. Given its isoelectric point, type A gelatin is positively charged below pH 7 and it is then expected to interact with G-C18:1 negative membranes.

In this work we have used a broad set of polyelectrolytes, of which the different chemical nature let us explore various aspects of their interactions with G-C18:1. If the nature of the polyelectrolyte (stiffness, charge density, …) is known to strongly affect the morphology and structure of PESCs,[14,32] in this work we show that: 1) when the *Co* and *MLWV* phases are formed, the structure of the corresponding colloidal structures is very similar, whichever the polyelectrolyte used, even if local phenomena like swelling or long-range order may vary from one polyelectrolyte to another. 2) The *Co* and *MLWV* phases are only obtained with polyelectrolytes with a net positive charge, that is polycations. 3) The *MLWV* phase is always preceded by the *Co* phase, which seems to be a necessary condition to drive the isostructural and isodimensional *Co*-to-*MLWV* transition. This phenomenon does not occur when gelatin is employed and where the *MLWV* phase is not observed. On the contrary, the *MLWV* phase is obtained for the CHL system, despite the fact that we do not have a proof by SAXS of the *Co* phase. In this regard, we must outline that the SAXS signal for the [G-C18:1 + CHL] system at basic pH is dominated by the precipitated CHL phase, which we think to be in major amount but not the only phase. Cryo-TEM shows the presence of an unknown fraction of complex coacervates, which we believe to be source of the *MLWV* phase at pH below 7. We also believe that the higher disorder of the *MLWV* phase in the [G-C18:1 + CHL] system (broader first order diffraction peak compared to the PLL-derived MLWV in Figure 4e) could be attributed to the smaller fraction of the initial *Co* phase. In other words, the presence of a less ordered *MLWV* phase in the CHL system could then the indirect proof that probably a small fraction of the *Co* phase forms in the CHL system.

**Quantitativity and size control**

If the synthesis of PESCs involving vesicles and polyelectrolytes, and eventually forming MLWV, has long been addressed in the literature,[37,68,69] very few studies, if none, address the issue of quantitativity in relationship to the mechanism of formation. In particular, the synthesis of MLWV from a continuous isostructural phase transition from a coacervate phase has not been addressed before, because MLWV are generally obtained by mixing vesicles and polyelectrolytes in solution.[19,29,32–35,37] If some authors state that the formation of MLWV is driven by the lipid:polyelectrolyte ratio, other authors show that a mix of agglutinated vesicles and MLWV are actually obtained.[38,39] Other procedures could probably be followed to increase this control when working with pre-formed vesicles, such as the insertion of the polymer into the hydrophobic vesicle bilayer, which was reported in the case of polycations bearing pendant hydrophobic groups.[37,70] However, it was found that such interaction could be accompanied by lateral lipid segregation, highly accelerated transmembrane migration of lipid molecules (polycation-induced flip-flop), incorporation of adsorbed polycations into vesicular membrane as well as aggregation and disruption of vesicles.[70]

To evaluate the amount of MLWV with respect to agglutinated vesicles, we compare the sample obtained by continuous *Co*-to-*MLWV* phase transition with a sample obtained by the more classical approach consisting in mixing G-C18:1 single-wall vesicles and polyelectrolyte, the main one employed in the literature of MLWV. If SAXS can prove the presence of a multilamellar structure, it cannot be easily employed to quantify and discriminate between the two structures. For this reason, instead of SAXS, we evaluate the content of MLWV between the two methods of preparation by combining cryo-TEM with optical microscopy using crossed polarizers. If cryo-TEM can differentiate between agglutination and MLWV, its high magnification is poorly compatible with good statistics, unless a large number of images are recorded. On the contrary, optical microscopy using cross polarizers is the ideal technique to differentiate, on the hundreds of micron scale, between MLWV and agglutinated vesicles: multilamellar structures (but not single-wall vesicles) show a characteristic maltese cross pattern[71] under crossed polarizers, found both in concentric lamellar emulsions[72] and in spherical lamellar structures.[73]

Cryo-TEM of samples obtained from a *Co*-to-*MLWV* phase transition was shown in Figure 4 and, as already commented above, they show a massive presence of vesicular structures having multilamellar walls, as also confirmed by the corresponding SAXS data presented in Figure 4. Figure 6 shows two representative microscopy images of a typical sample prepared with the same approach; images are collected under white (a,d) and polarized light with polarizers at 0°-90° (b,e) and 45°-135° (c,f). The system is characterized by a large number of vesicles highly heterogeneous in size but all below ~10 μm. Under polarized light and crossed polarizers, the entire material displays a typical maltese cross colocalized with each vesicle. Despite the aggregation of the vesicles, also observed with cryo-TEM, maltese crosses are well-defined and nicely separated and each identifying single multilamellar wall vesicles. The entire material displays such a characteristic birefringency, strongly suggesting a quantitative presence of MLWV.

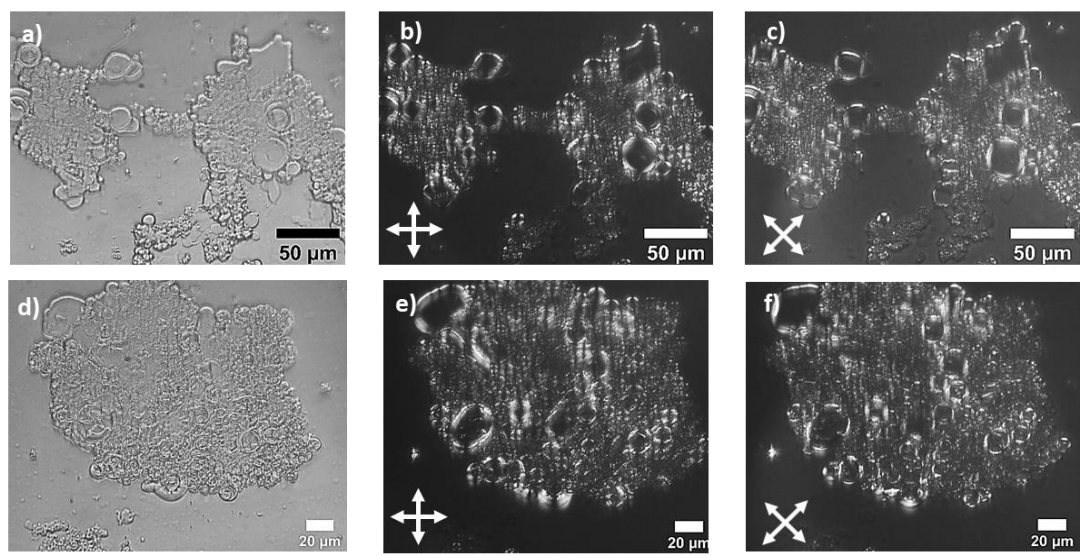

**Figure 6 – Optical microscopy images recorded on a [G-C18:1 + PLL] solution ($C_{\text{G-C18:1}}$= $C_{\text{PLL}}$= 2.5 mg.mL$^{-1}$) at pH 3.9 obtained from a *Co*-to-*MLWV* phase transition. a,d) white light and polarized light with cross polarizers set at b,e) 0-90° and c,f) 45-135°.**

The experiment consisting in mixing acidic solutions (pH 3.8) of pre-formed G-C18:1 single-wall vesicles and PLL is shown in Figure 7. A preliminary investigation by optical microscopy results in a different behavior and distribution of signal with respect to the sample obtained through the *Co*-to-*MLWV* phase transition. Figure 7a shows representative images of a sample being constituted of aggregated objects, each of size below 1 μm, expected for G-C18:1 vesicles.[43] The corresponding images recorded using crossed polarizers (Figure 7b,d) show a broad, undefined, birefringency associated to the aggregates with little, if no, content of maltese crosses. The featureless, generalized, birefringency signal suggests that MLWV are either not formed or they form in small amounts, in good agreement with the data presented by others.[38,39] This assumption is confirmed by cryo-TEM images recorded on the same system and showing a mixture of structures including agglutinated vesicles but also "cabbage-like" and multilamellar structures (Figure 7e-f).

The massive presence of MLWV structures obtained through the phase transition process compared to the mixture of structure obtained from a direct mixing of pre-formed vesicles-polyelectrolyte solutions confirms the crucial role of the complex coacervates in the formation of MLWV: coacervation seems to be a requirement to the extensive formation of vesicular structures with multilamellar walls.[41] This is also in agreement with the data obtained from the [G-C18:1 + gelatin] system presented in Figure 5 and prepared using the pH variation approach. Also in that case, the absence of a complex coacervate phase had as a consequence the absence of the *MLWV* phase. An additional piece of evidence comes from the CHL system, in which the limited amount of the *Co* phase generates a more disordered *MLWV* phase. Combination of the data obtained with gelatin and employing the *in situ* pH variation with the data obtained by mixing vesicle and polyelectrolyte solutions at a given pH demonstrates the importance of the precursor *Co* phase during the phase change method in order to obtain a massive presence of MLWV structures.

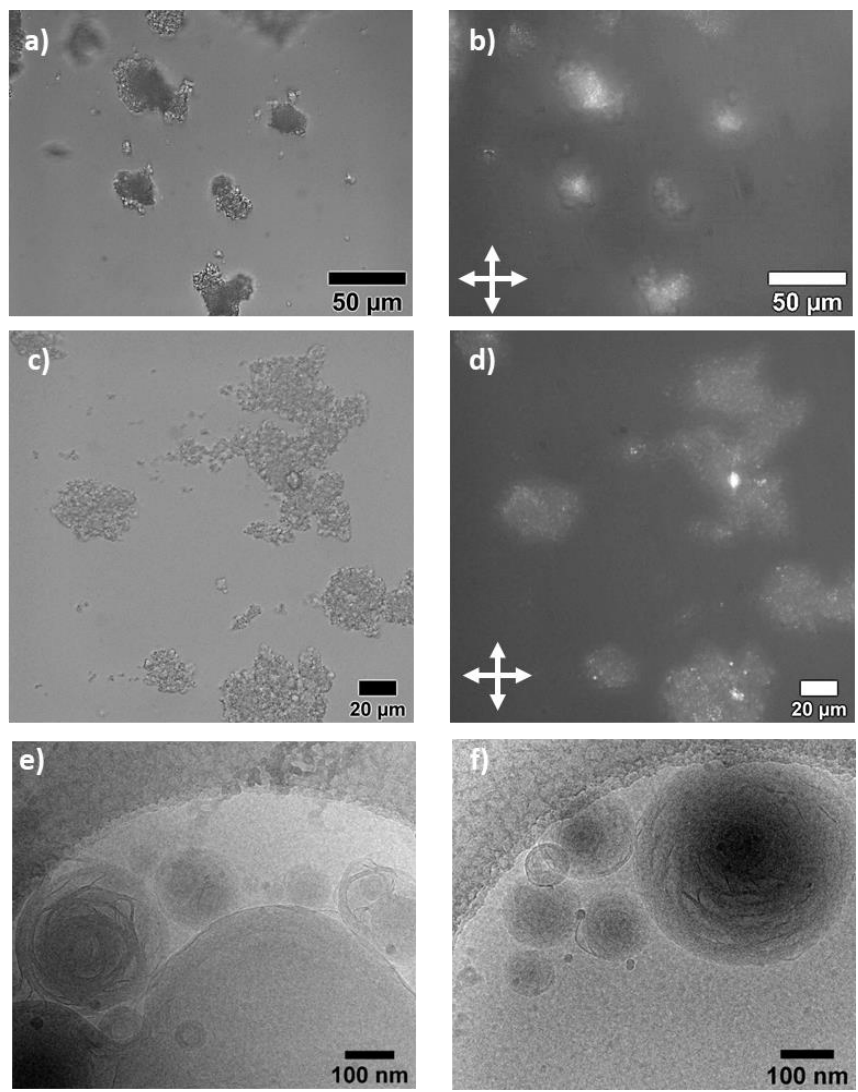

**Figure 7 – a-d) Optical microscopy images recorded on a mixture of [G-C18:1] single-wall vesicles and [PLL] solutions ($C_{\text{G-C18:1}}$= $C_{\text{PLL}}$= 2.5 mg.mL$^{-1}$) both prepared at pH= 3.8. a,c) white light and b,d) polarized light with cross polarizers set at 0-90°. Images in e,f) are recorded on the same sample by mean of cryo-TEM.**

If the *Co*-to-*MLWV* phase transition is able to quantitatively produce MLWV, its main drawback is the poor control over their size distribution, as shown both by TEM and optical microscopy. To improve this point, we employed filtration (Figure 8a-c) and sonication (Figure 8d-f), these methods being known to efficiently control vesicles size distribution,[74] but unclear whether or not they have any deleterious impact on the MLWV structure. According to the cryo-TEM data in Figure 8a-c, filtration (pore size, $\varphi$= 450 nm) promotes the stabilization of colloidally-stable spherical MLWV, of which the diameter seems to be contained between 50 nm and about 300 nm, in agreement with the filter pore size. Concerning the effect of sonication, Figure 8d-f also shows a large number of spherical, un-aggregated, MLWV colloids, although the diameter appears to be bigger of several hundred nanometers if compared to the filtered sample. The cryo-TEM results are confirmed by intensity-filtered DLS experiments, presented in Figure 8g. The as-prepared sample (black curve) shows a MLWV distribution centered at 716 nm, while the filtered sample shows a distribution centered at 460 nm. To better evaluate the impact of sonication, we tested the influence of sonication time and according to DLS data (Figure 8g) we find that at $t$= 30' the size distribution is centered at higher diameter values and it is even broader than the as-prepared sample. Applying the same sonication conditions, but over a longer period of time ($t$= 1' or $t$= 1'30''), it is possible to reduce the MLWV diameter even if the size distribution is broader than the filtration approach, in agreement with the cryo-TEM data.

These experiments show that control of the size distribution of MLWV is possible using standard methods employed in liposome science without perturbing the multilamellar wall structure.

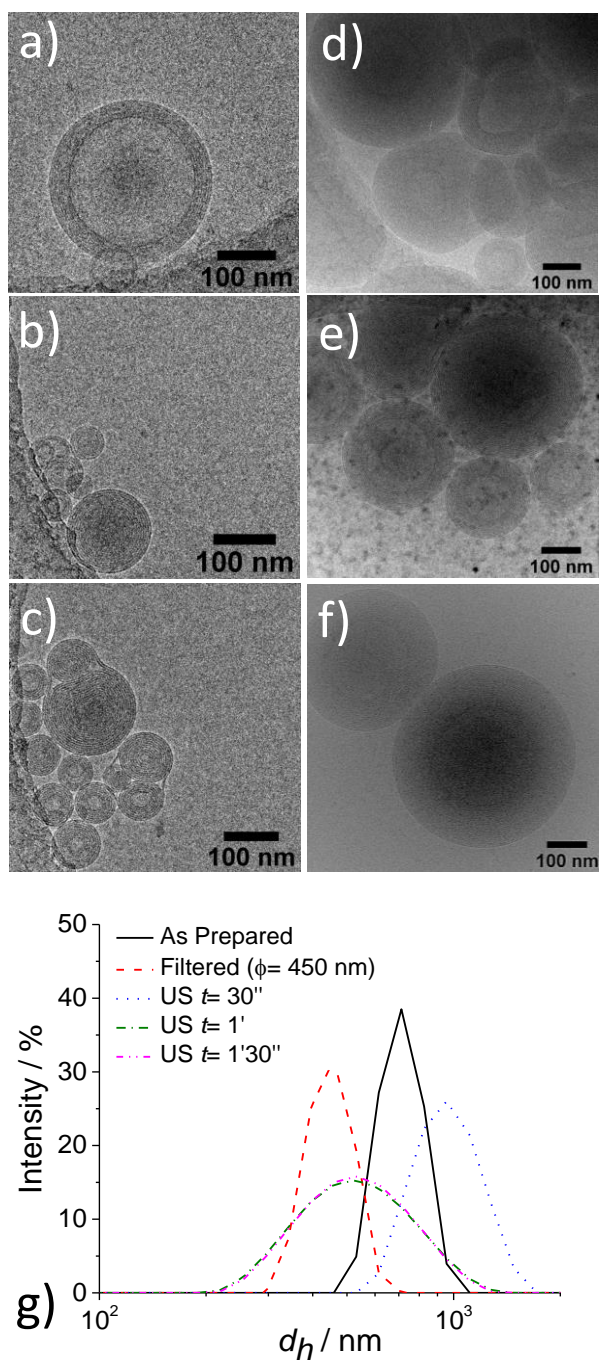

**Figure 8 - Cryo-TEM images of a [G-C18:1 + PLL] solution (pH=5, $C_{\text{G-C18:1}}$= $C_{\text{PLL}}$= 2.5 mg.mL$^{-1}$) prepared using the *Co*-to-*MLWV* approach and a-c) filtered through a $\varphi$= 450 nm pores membrane or d-f) sonicated (ultrasound, US, technical data: $t$= 1', $P$= 40 W, Ampl.= 40%, freq.= 100%). g) profiles of the as prepared (black curve), filtered ($\varphi$= 450 nm pores membranes, red curve) and sonicated (US, technical data: $P$= 40 W, Ampl.= 40%, freq.= 100%, time is given on graph) MLWV samples**

**Conclusion**

This work addresses the synthesis of multilamellar wall vesicles (MLWV) using a recently developed method involving a pH-stimulated transition from a complex coacervate phase (*Co*) instead of a polyelectrolyte-driven vesicle agglutination, classically-employed in the preparation of MLWV polyelectrolyte surfactant complexes (PESCs). We use a combination of a stimuli-responsive microbial glycolipid biosurfactant and a polyelectrolyte, mainly polyamines (either synthetic or natural). The deacetylated acidic C18:1 glucolipid, G-C18:1, undergoes a micelle-to-vesicle phase transition from alkaline to acidic pH. In the alkaline pH domain, its phase behavior is mainly characterized by negatively-charged micelles. In the presence of a positively charged polyelectrolyte, G-C18:1 forms a *Co* phase. Upon lowering pH below the micelle-vesicle boundary, *in situ* SAXS experiments show a continuous isostructural and isodimensional transition between the *Co* and *MLWV* phase. The acidification process reducing the negative charge density, the micellar aggregates embedded in the *Co* phase undergo a decrease in the local curvature, which drives the transition from spheres to membranes, made of interdigitated G-C18:1 molecules. The membrane has a residual negative charge density, responsible for the strong electrostatic interaction with the polyelectrolyte, crucial to maintain the membranes together. At lower pH, the membrane charge density becomes low and interactions with the polyelectrolyte decrease. This phenomenon promotes intra-chain electrostatic repulsion interactions and eventually encourage the lamellar region to swell. Finally, when the membrane reaches neutrality, polymeric repulsion becomes strong enough to disassemble the lamellae. The polyelectrolyte will most likely be entirely solvated and at sufficiently low pH (< 3) the G-C18:1 precipitates in the form of a lamellar phase, possibly free of the polyelectrolyte, a behavior characteristic of the control lipid solution at the same pH.

We employ four polyelectrolytes, synthetic and natural and with different characteristics of rigidity and charge density (chitosan, poly-L-Lysine, polyethylene imine, polyallylamine); however, the nature of the polyelectrolyte does not seem to be a relevant parameter concerning the fate of the transition, as otherwise found for most PESCs. This may be explained by the strong proximity between the lipid and the polyelectrolyte throughout the isostructural *Co*-to-*MLWV* transition. If the method described in this work does not allow a tight control over the size distribution of MLWV, we also find that the multilamellar wall structure is stable against filtration and sonication, two common methods employed to control the size of vesicles. Last but not least, we show that if we employ the classical approach consisting in mixing pre-formed vesicles with a cationic polyelectrolyte solution at a given pH, we find a much broader structural diversity, including agglutinated single-wall vesicles, multilamellar but also cabbage-like structures, in agreement with previous literature studies.

All in all, this work establishes the ground for the preparation of a new generation of fully biobased, stimuli-responsive, PESCs, of which the potential fields of applications could span from cosmetics to home-care products.

**Acknowledgements**

Diamond synchrotron radiation facility (U. K.) is acknowledged for accessing to the B21 beamline and financial support (proposal N° 23247), as well as Soleil Synchrotron facility for accessing the Swing beamline and financial support (proposal N° 20190961). Ghazi Ben Messaoud (DWI-Leibniz Institute for Interactive Materials, Aachen, Germany) is kindly acknowledged for helpful discussions. We thank Dr. S. Roelants, Prof. W. Soetaert and Prof. C. V. Stevens at Gent University for providing us the glycolipid. Sorbonne Université (contract N°3083/2018) is acknowledged for financial support of CS. Authors kindly acknowledge the French ANR, Project N° SELFAMPHI - 19-CE43-0012-01.

**Authors' contributions**

CS performed the experiments, analyzed the data and wrote the manuscript. PG performed the cryo-TEM experiments. NC and JP assisted the SAXS experiments. NB conceived and supervised the work and wrote the manuscript.

**Supporting Information**: Figure S1

# Supplementary information

**Synthesis of multilamellar walls vesicles polyelectrolyte-surfactant complexes from pH-stimulated phase transition using microbial biosurfactants**

**Chloé Seyrig[a], Patrick Le Griel[a], Nathan Cowieson,[b] Javier Perez,[c] Niki Baccile[a]**

[a] Sorbonne Université, Centre National de la Recherche Scientifique, Laboratoire de Chimie de la Matière Condensée de Paris , LCMCP, F-75005 Paris, France

[b] Diamond Light Source Ltd, Diamond House, Harwell Science & Innovation Campus, Didcot, Oxfordshire, OX11 0DE

[c] Synchrotron SOLEIL, L'Orme des Merisiers Saint-Aubin, BP 48 91192 Gif-sur-Yvette Cedex

*Content* : Figure S1, Figure S2

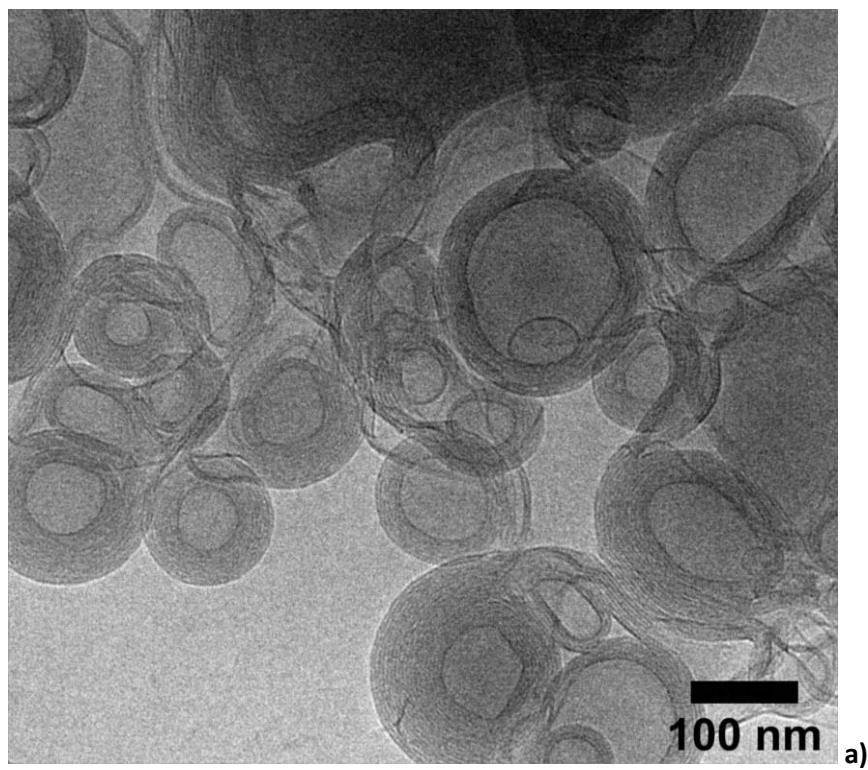

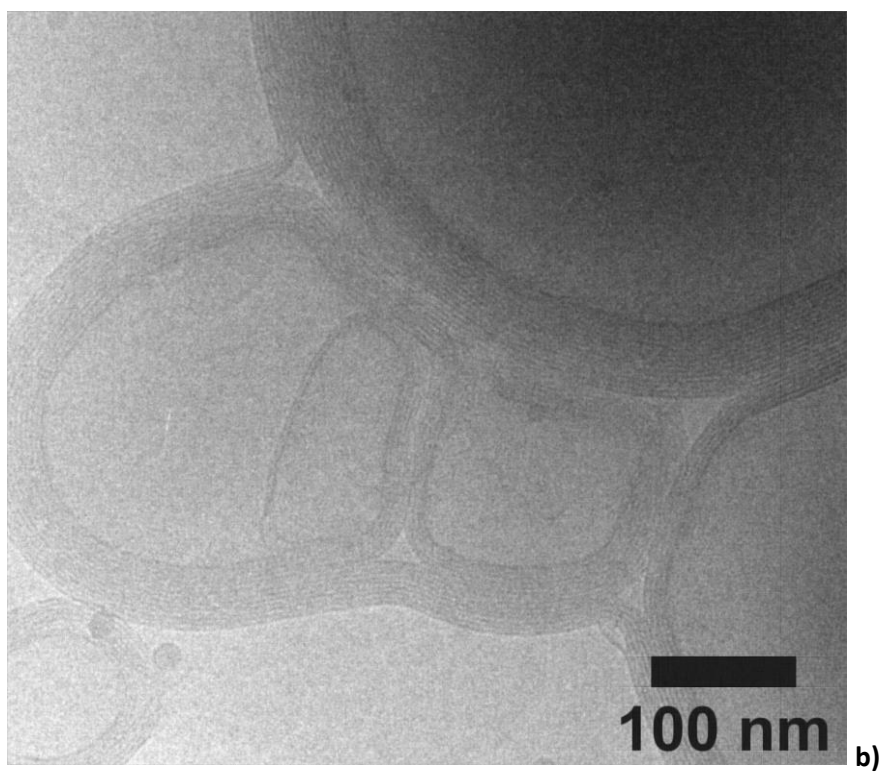

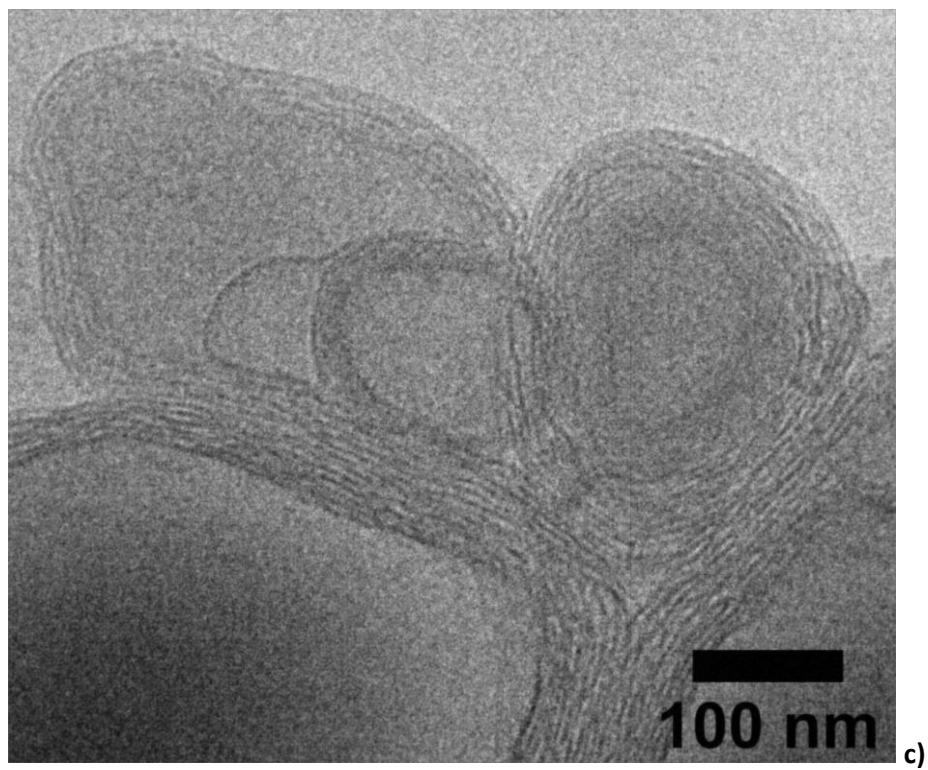

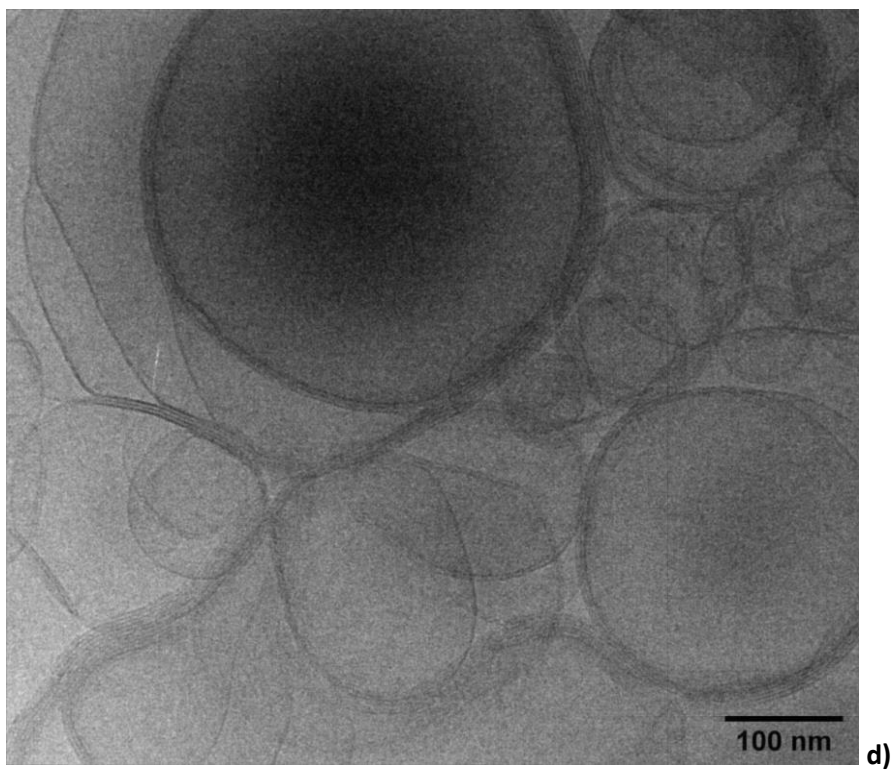

**Figure S 1 -** Additional cryo-TEM images and zooms on layers of *MLWV* made of G-C18:1 (2.5 mg.mL$^{-1}$) + a) CHL (1 mg.mL$^{-1}$, pH 4.87), b) PAH (0.25 mg.mL$^{-1}$, pH 4.25), c) PEI (2.5 mg.mL$^{-1}$, pH 5.33) and d) PLL (2.5 mg.mL$^{-1}$, pH 4.70)

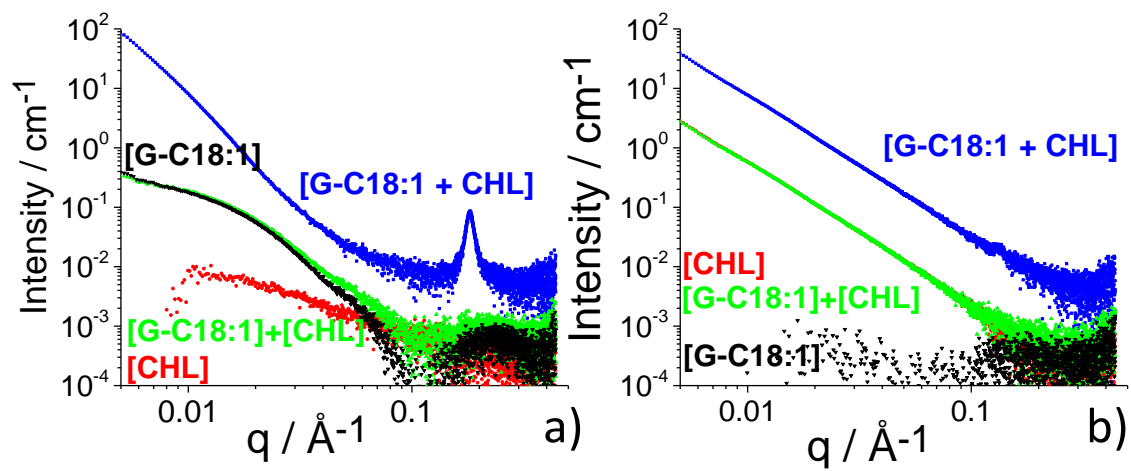

**Figure S 2** - SAXS profiles of [G-C18:1] (black) and [CHL] (red) control and [G-C18:1 + CHL] (blue) solutions ($C_{\text{G-C18:1}}$= 2.5 mg·mL$^{-1}$, $C_{\text{CHL}}$= 1 mg·mL$^{-1}$) at a) pH= 4.73 and b) pH= 8.81. The green [G-C18:1] + [CHL] profiles correspond to the arithmetic sum of [G-C18:1] + [CHL] individual SAXS profiles.